%% file: main.tex
\documentclass[acmsmall]{acmart}

\setcopyright{none}
\AtBeginDocument{%
  \providecommand\BibTeX{{%
    \normalfont B\kern-0.5em{\scshape i\kern-0.25em b}\kern-0.8em\TeX}}}

  \usepackage{multirow}
  \usepackage{diagbox}
  \usepackage{graphicx}
  \usepackage{color}
  
  \usepackage{amsmath, amssymb, bm}
  \usepackage{algorithm}
\usepackage{algorithmic}
\floatname{algorithm}{Algorithm}

\setcopyright{acmcopyright}
\copyrightyear{2023}
\acmYear{2023}
\acmDOI{10.1145/1122445.1122456}

\acmJournal{TOIS}
\acmVolume{37}
\acmNumber{4}
\acmArticle{111}
\acmMonth{8}

\begin{document}

\title{PEPT: Expert Finding Meets Personalized Pre-training}


\author{Qiyao Peng}
\email{qypeng@tju.edu.cn}
\affiliation{
\streetaddress{School of New Media and Communication}
  \institution{Tianjin University}
  \city{Tianjin China}
}
\author{Hongyan Xu}
\email{hongyanxu@tju.edu.cn}
\affiliation{
\streetaddress{College of Intelligence and Computing}
  \institution{Tianjin University}
  \city{Tianjin China}
}
\author{Yinghui Wang}
\email{wangyinghui@tju.edu.cn}
\affiliation{
\streetaddress{Key Laboratory of Information System and Technology}
  \institution{Beijing Institute of Control and Electronic Technology}
  \city{Beijing China}
}
\author{Hongtao Liu}
\email{htliu@tju.edu.cn}
\affiliation{
  \institution{Du Xiaoman Technology}
  \city{Beijing China}
}
\author{Cuiying Huo}
\email{huocuiying@tju.edu.cn}
\affiliation{
\streetaddress{College of Intelligence and Computing}
  \institution{Tianjin University}
  \city{Tianjin China}
}
\author{Wenjun Wang}
\email{wjwang@tju.edu.cn}\thanks{*Qiyao Peng and Hongyan Xu are equal contributions. Wenjun Wang is the corresponding author}
\affiliation{
\streetaddress{College of Intelligence and Computing}
  \institution{Tianjin University}
  \city{Tianjin China}
}
\affiliation{
\streetaddress{He is also in Yazhou Bay Innovation Institute}
  \institution{Hainan Tropical Ocean University}
  \city{Hainan China}
}

\renewcommand{\shortauthors}
{Qiyao Peng et al.}

\input{0-abstract}



\maketitle

\input{1-introduction}
\input{2-relatedworks}
\input{3-method}
\input{4-experiments}
\input{5-results}
\input{6-conclusion}

\begin{acks}
This work was supported by: (1) Hainan Province Science and Technology Special Fund, Grant No: ZDYF2024SHFZ051; (2) The Research Program of Yazhou Bay Innovation Research Institute, Hainan Tropical Ocean University, Grant No: 2023CXYZD001.
\end{acks}

\bibliographystyle{ACM-Reference-Format}
\bibliography{trans}

\end{document}

%% file: 0-abstract.tex
\begin{abstract}
Finding experts is essential in Community Question Answering (CQA) platforms as it enables the effective routing of questions to potential users who can provide relevant answers.
The key is to personalized learning expert representations based on their historical answered questions, and accurately matching them with target questions.
Recently, the application of Pre-trained Language Models (PLMs) have gained significant attraction due to their impressive capability to comprehend textual data, and are widespread used across various domains.
There have been some preliminary works exploring the usability of PLMs in expert finding, such as pre-training expert or question representations.
However, these models usually learn pure text representations of experts from histories, disregarding personalized and fine-grained expert modeling.
For alleviating this, we present a personalized pre-training and fine-tuning paradigm, which could effectively learn expert interest and expertise simultaneously.
Specifically, in our pre-training framework, we integrate historical answered questions of one expert with one target question, and regard it as a candidate aware expert-level input unit.
Then, we fuse expert IDs into the pre-training for guiding the model to model personalized expert representations, which can help capture the unique characteristics and expertise of each individual expert.
Additionally, in our pre-training task, we design: 1) a question-level masked language model task to learn the relatedness between histories, enabling the modeling of question-level expert interest; 2) a vote-oriented task to capture question-level expert expertise by predicting the vote score the expert would receive.
Through our pre-training framework and tasks, our approach could holistically learn expert representations including interests and expertise.
Our method has been extensively evaluated on six real-world CQA datasets, and the experimental results consistently demonstrate the superiority of our approach over competitive baseline methods. \footnote{This work is an extended and revised version of research that published in EMNLP~2022~\cite{peng2022expertplm}.}
\end{abstract}

%% file: 1-introduction.tex
\section{INTRODUCTION}

Recently, the popularity of online CQA websites based on user-generated content has been continuously increasing~\cite{zhu2011ksem,srba2016comprehensive,amendola2024understanding}.
For instance, some large websites such as Zhihu~\footnote{https://www.zhihu.com/}, StackExchange~\footnote{https://stackexchange.com/}, Quora~\footnote{https://www.quora.com/} provide convenient bridges to exchange knowledge and have successfully garnered a substantial user base~\cite{HuQFX19,HuQFX21}.
These websites allow users to ask questions or provide responses related to topics of their interest or expertise~\cite{fang2016community,peng2022towards}.
Due to the large participation, some questions remain unanswered for extended periods, resulting in unsatisfactory response times for askers~\cite{yuan2020expert}.
Therefore, routing questions to suitable experts, who are capable of providing satisfactory answers, is crucial to CQA websites~\cite{chang2013routing}.
Expert finding on the CQA websites can effectively recommend questions to suitable experts~\cite{amendola2024leveraging,amendola2024towards} capable of providing high-quality answers.
Then, the askers could receive answers quickly~\cite{liu2022cikm}. 
This approach has recently garnered considerable attention in both academic and industry.

A key challenge in expert finding is the accurate matching of candidate experts with target questions~\cite{li2019personalized,zhang2020temporal}.
Thus, it is crucial to learn 
personalized \textbf{expert representations} to achieve precise question-expert matching.
Existing expert finding methods mostly learn expert representations from limited supervised data (e.g., experts' previously answered questions) and subsequently compute the expert-question matching scores.
For example,
Pal et al.~\cite{pal2011early} model users from their early historical activities, then use classification and ranking
models to find potential experts.
NeRank~\cite{li2019personalized} considers the question raiser's profile, question content and question answerer ID embedding to learn entity representations with a network embedding method, which could help capture personalized expert representations.
Peng et al.~\cite{peng2022towards} design a multi-view learning paradigm to learn multi-view question representations based on the question title, body and tag information. Then the model introduces the expert ID information to personalized compute question-expert relatedness for identifying experts.
     
In recent years, the language modeling capability of PLMs (e.g., BERT~\cite{devlin2018bert}), has demonstrated remarkable improvements.
Pre-training the Transformer Encoder model~\cite{vaswani2017attention}, on large unsupervised corpora enables models to learn more generalized knowledge representations.
Extensive research has shown that pre-trained models are capable of capturing intricate sequence patterns from large amounts of raw data.
These models can then transfer this knowledge to a range of downstream NLP tasks, especially in situations where limited training data is available~\cite{brown2020language,raffel2020exploring}.
Inspired by this, researchers have been exploring the feasibility of pre-training technology for finding suitable experts~\cite{liu2022expertbert,peng2022expertplm}.
For example, one preliminary work, ExpertPLM~\cite{peng2022expertplm} designs a tailored pre-training model for modeling expert based on the historical answered questions, which could help learn expert representations from unsupervised data.

Although previous studies have demonstrated effectiveness, we argue that they still have several limitations that need to be addressed:

\textbf{1) Interest Modeling}. Experts' interests can reflect their willingness to answer questions, which could be captured from their historical answered questions. Existing Pre-trained Language Model (PLM)-based expert finding methods typically concatenate all the questions an expert has answered. However, this approach only captures experts' interests based on word-level information and neglects question-level interest. In fact, expert interest is better represented by question-level semantics rather than individual keywords. Therefore, how to capture the question-level interest is important during pre-training.

\textbf{2) Expertise Modeling}. In fact, the ability of an expert to answer a target question plays a crucial role in identifying a suitable expert. For instance, one expert has answered questions ``Was public transportation free in the Soviet Union?'' and ``Is there any record of ancient Israelite contact with Indo-Europeans?'', which could reflect his interest in ``Soviet Union'' and ``Indo-Europeans''. However, the vote scores received for two different questions can vary significantly (e.g., $+26$ and $-5$), indicating variations in the expert's expertise across different questions. Nevertheless, most existing pre-trained language model-based methods struggle to capture an expert's ability to answer a specific question.

\textbf{3) Personalized Modeling}. It is intuitive that different experts exhibit diverse personalized characteristics. For example, there is one question ``Why did Hitler ... Soviet Union ... fighting the United Kingdom?'' waiting for answers. If one expert is more concerned about ``Hitler'' than ``Britain'', and another expert is more concerned about ``Britain'' than ``Hitler'', their concerns on the question are different. Hence, the question differs in usefulness for them when learning their representations. However, existing PLM-based methods usually ignore these essential personalized characteristics.

Hence, for alleviating the above limitations, we propose to enhance the expert finding via a \textbf{P}ersonalized \textbf{E}xpert-level \textbf{P}re-\textbf{T}raining language model (named \textbf{PEPT}), which could personalized pre-train expert representations before the downstream task expert finding.
Our approach addresses the aforementioned limitations as follows:

1) Interest modeling. To enhance the pre-trained model's understanding of experts' interests, we integrate the expert's historical answered questions as a source of expert-level interest information and utilize it as input for the model. Furthermore, we design a question-level pre-training strategy to capture underlying and inherent relationships between different historical questions. Compared to the traditional masked language model task for pre-training expert interests, our approach could learn more fine-grained expert interest features.

2) Expertise modeling. One way to model an expert's ability to answer questions is by using the vote scores, which indicate the level of CQA user appreciation for a particular answer. In other words, higher vote scores generally indicate higher levels of expertise associated with the user providing that answer. Thereby, to learn an expert's fine-grained abilities of answering questions, we incorporate his/her historical answered question input embedding with corresponding encoded vote scores. Additionally, given one target question, we introduce a vote-oriented task to predict the vote score an expert would receive, empowering the model a better understanding of an expert's ability to answer the target question.

3) Personalized modeling. To further enhance personalized expert characteristics learning, we introduce the unique Expert ID into the model, which could indicate different expert personalized characteristics. One simplistic approach is to translate numerical expert IDs into their corresponding word embeddings, but this method can lead to excessive vocabulary size. Inspired by continuous/soft prompts~\cite{lester2021power}, we directly input Expert ID embeddings into the model by incorporating them with the expert-level information, which enhances the model to achieve personalized pre-training.

In a word, we propose to unify heterogeneous information into the pre-training framework to provide rich information for learning expert representation. 
Furthermore, we design a multi-task pre-training paradigm for guiding the model to capture fine-grained expert interest and expertise.

During finding experts, we construct the model input like that in pre-training for ensure the input consistency, then compute the expert-question matching score.
This approach enables us to directly identify suitable experts through a fine-tuning way.

Our proposed model has the following main contributions:
\begin{itemize}
    \item We design a personalized pre-training language model paradigm in expert finding, named \textbf{PEPT}, aiming to pre-train personalized and fine-grained expert representations, and then conduct the expert finding in a fine-tuning way. To our knowledge, this is the first study to design a personalized pre-training paradigm for expert finding, enabling to capture more comprehensive expert features derived from unsupervised data.
    
    \item We unify the historical questions, corresponding vote scores, and personalized information in the pre-training stage to model experts. Beneficial from the designing of this, the expert finding can be directly conducted based on the pre-trained model. Furthermore, during pre-training, we design a question-level MLM task and a vote-oriented task, which allows to better model fine-grained expert characteristics.
    
    \item Through experiments on six real-world community question answering website datasets, we can find PEPT outperforms the recent baselines. These results demonstrate the superiority of PEPT in enhancing expert finding. Additionally, our further analysis showcases the effectiveness of different model modules and the potential of pre-trained representations for improving expert finding.
\end{itemize}

\textbf{Differences between this work and the conference version.}
This work is an extended and revised version of research that appeared at EMNLP~2022~\cite{peng2022expertplm}.
The current version differs significantly from its preliminary conference version in the pre-training paradigm.
Specifically,
\begin{enumerate}
    \item We introduce the personalized information of experts into the expert pre-training, which enables the model to learn the unique representations of experts effectively. However, the previous version did not take into account such personalization;
    \item This work designs a more fine-grained expert pre-training architecture compared with our previous version.
    Specifically, our conference version only considers modeling the coarse-grained expert-level characteristics, and our extended version designs two question-level pre-training tasks, which could empower the model ability to capture more fine-grained interests and capabilities;
    \item We integrate target questions in the expert-level information, which can help the model pre-train candidate-aware expert representations. 
    Another advantage of this design is that it can eliminate the gap between the pre-training and expert finding, i.e., there is no need to design additional encoders during fine-tuning like the previous version but can be unified in one model;
    \item In addition, this paper conduct various of experiments aimed at corroborating the effectiveness of our pre-training paradigm.
\end{enumerate}

%% file: 2-relatedworks.tex
\section{RELATED WORK}

We provide a concise literature review on three key topics relevant to our proposed method: Pre-training for Natural Language Processing (NLP), Pre-training for Recommender Systems (RS), and Expert Finding in this section.

\begin{table*}
\centering
\caption{Characteristics of Different Expert Finding Methods. The model possesses the feature marked with a check mark, while the feature marked with a cross is not possessed by the model.}
\resizebox{0.7\textwidth}{!}{
\begin{tabular}{c|c|c|c|c}
\toprule[1pt]
\diagbox{Method}{Characteristic} & Interest & Expertise & Personalized & Pre-training  \\
\midrule
CNTN~\cite{qiu2015convolutional} & $\times$ & $\times$ & $\times$ & $\times$   \\
\midrule
NeRank~\cite{li2019personalized} & $\surd$ & $\times$ & $\surd$ & $\times$  \\
\midrule
TCQR~\cite{zhang2020temporal} & $\surd$ & $\times$ & $\surd$ & $\times$  \\
\midrule
RMRN~\cite{fu2020recurrent} & $\surd$ & $\times$ &  $\times$ & $\times$  \\
\midrule
UserEmb~\cite{ghasemi2021user} & $\surd$ & $\times$ &  $\surd$ & $\times$  \\
\midrule
PMEF~\cite{peng2022towards} & $\surd$ & $\times$ & $\surd$ & $\times$   \\
\midrule
EFHM~\cite{peng2022towardskbs} & $\surd$ & $\times$ & $\surd$ & $\times$   \\
\midrule
ExpertBert~\cite{liu2022expertbert} & $\surd$ & $\times$ &  $\times$ & $\surd$  \\
\midrule
\textbf{PEPT} & \textbf{$\surd$} & \textbf{$\surd$} &  \textbf{$\surd$} & \textbf{$\surd$}  \\
\bottomrule[1pt]
\end{tabular}
}
\label{tab:methods}
\end{table*}

\subsection{Pre-training for Natural Language Processing}

Recently, various of Pre-Training Models (PTMs) have demonstrated the efficacy of pre-training on large corpora and the capability to acquire general language knowledge, which can be effectively transferred to various downstream NLP tasks.~\cite{yang2020alternating,yang2023ganlm,bai2023qwen}.
Earlier word embedding methods, such as Word2Vec , Glove learned good word embeddings by capturing word co-occurrence information. 
These embeddings derived from these models could significantly improve various NLP missions (e.g., text classification, named entity recognition) by capturing the semantic meanings of words.
However, these models are context-independent and predominantly trained using shallow models, which may restrict their effectiveness in specific scenarios that necessitate a more profound comprehension of language context.

To overcome these limitations, various of pre-training methods~\cite{brown2020language,wang2023rolellm,sun2024unicoder} based on the Transformer architecture~\cite{vaswani2017attention} have been developed, showing their significant improvements in learning contextual language representations.
These approaches, such as BERT~\cite{devlin2018bert}, MASS~\cite{mass2019}, and BART~\cite{lewis2020bart}, mainly pre-train model parameters based on various neural network (e.g., transformer encoder, transformer decoder) and language modeling objectives.
Then fine-tune the model parameters with downstream supervised data.
MASS~\cite{mass2019} is a transformer-based model that introduces the ``masked target sequence reconstruction'' task to address the limitation of traditional sequence-to-sequence models in machine translation.
By employing pre-training and fine-tuning paradigms, it can proficiently utilize extensive corpora to acquire comprehensive language knowledge.
At the same time, the model could integrate task-specific information during the fine-tuning phase.
Thus, these models have the potential to enhance the performance of diverse downstream tasks.

However, most existing PLMs are carefully designed at the corpus-level to capture universal language knowledge, limiting their ability to model experts. 
Therefore, it is nontrivial to directly employ the original pre-training models and the CQA corpus (e.g., question titles) for learning expert representations.

\subsection{Pre-training for Recommender Systems}

In recent times, pre-training technology has been increasingly used in recommendation models to enhance the performance of recommendation systems.
Specifically, these methods~\cite{KangM18,sun2019bert4rec} leverage pre-training techniques to learn item co-occurrence information, which can be applied to improve the recommendation.
For example, 
SASRec~\cite{KangM18} employed a self-attention based sequential model to identify which items
are ``relevant'' from a user’s action history for improving sequential recommendation.
Sun et al.~\cite{sun2019bert4rec} learned item representations via fusing both the left and right context, then utilized the cloze objective to train the model.
Cheng et al.~\cite{ChengYLXC21} proposed CLUE. It employed contrastive learning as its optimization objective during pre-training, with a focus on optimizing sequence-level representations.
UniSRec~\cite{HouMZLDW22} captured more transferable representations from the item texts for modeling users and items, which could effectively generalize to new recommendation scenarios.
Then it fine-tuned the pre-trained model for conducting recommendation and user-oriented downstream tasks, which obtained better performance.
Furthermore, many researchers have aimed to capture universal user embeddings that could be applied to various user-oriented downstream tasks~\cite{he2020para,xiao2021uprec}.
For example, 
PeterRec~\cite{he2020para} proposed a self-supervised model to learn user representations, which could be used to infer user profiles, such as the gender, age, preferences, etc. 
Xiao et al.~\cite{xiao2021uprec} proposed UPRec, which integrated heterogeneous user information and employed two pre-training tasks to learn user-aware representations.

However, unlike the above sequential recommendation tasks, expert finding on CQA websites is always faced with the challenge of cold-start (i.e., the target question is always a new question).
Moreover, it has unique characteristics that require specialized approaches, such as expertise modeling of experts, which aims to reflect the ability of experts to provide accurate and helpful answers. 
As a result, developing effective pre-training approaches for expert finding requires careful consideration of these unique characteristics.

\subsection{Expert Finding}

The primary objective of expert finding in the context of CQA websites is to locate individuals who possess the necessary expertise and knowledge to provide satisfactory responses to user questions~\cite{zhu2011cikm,yuan2020expert}.
The existing body of research in this domain can be broadly classified into two major categories: traditional expert finding methods and expert finding methods based on deep learning techniques.

Traditional methods often relied on feature engineering~\cite{robertson2009probabilistic,cao2012approaches,pal2012exploring} or topic modeling techniques~\cite{riahi2012finding,ji2013learning,yang2013cqarank} to learn question and expert representations.
Afterwards, the model utilized the question and expert representations for computing the relevance scores and routing questions to suitable experts.
For instance,
Zhou et al.~\cite{zhou2012classification} captured local-/global-level characteristics based on questions, users, and relations between them, then these features were fed to the SVM~\cite{smola2004tutorial} for finding suitable experts.
Chang et al.~\cite{chang2013routing} employed LDA~\cite{blei2003latent} to model the topic information from the question contents and took that information as the question representations for recommending experts.
Neshati et al.~\cite{NESHATI20171026} introduced two learning algorithms that considered the temporal-aware aspects of expert finding. They combine factors such as topic similarity to predict a user's likelihood of becoming an expert.
Deep learning-based methods~\cite{huang2013learning,qiu2015convolutional,wang2015long} leveraged neural networks to model questions and experts, then computed the question-expert relevance for finding experts~\cite{li2019personalized,fu2020recurrent,ghasemi2021user}.
For example, TCQR~\cite{zhang2020temporal} employed the pre-trained BERT to learn question semantic information and captured the answerer’s features based on the question semantic, temporal information and answerer  embedding for expert finding.
Qian et al.~\cite{qian2022heterogeneous} designed  a multi-perspective metapath-based network and considered vote scores for question routing, but they omitted the temporal interest of the expert.
HQExpert employed~\cite{liu2022high} different granularity semantic information and temporal interest to accurately identify experts and improve accuracy, but they could not consider personalized characteristics of the expert and the varying expert-question relationships.
Vaibhav et al.~\cite{krishna2022question} proposed a personalized expert finding approach to consider the similarity between the content of the recent activity of a user (the “domain expertise”) and the content of a given new question.
MPQR~\cite{qian2022heterogeneous} utilized a heterogeneous information network (HIN) constructed from answering records and voting information to personalized learn multi-perspective representations of question answerers, question raisers, and questions.
EFHM~\cite{peng2022towardskbs} is a multi-grained (word-level, question-level and expert-level) hierarchical matching framework for expert finding in CQA websites, which could capture more fine-grained matching relationships and achieve superior performance. In the expert-level, the model introduce the expert ID embedding to personalized learn expert representations.

In recent years, PLMs have emerged as a powerful tool for improving the performance of various NLP tasks, including textual similarity, etc.~\cite{ijcai2021p462,xu2023group}.
This is primarily due to their superior ability to understand and analyze complex textual data, enabling them to capture more nuanced features and relationships within language.
Some works have focused on leveraging the capabilities of PLMs to develop advanced expert finding techniques~\cite{liu2022expertbert,peng2022expertplm}.
For instance, Liu et al.~\cite{liu2022expertbert} designed a semi-supervised pre-training paradigm for expert finding, which could incorporate downstream tasks into the pre-training and improve the model's ability to utilize limited data.

Based on the existing methods discussed above, we have identified several key characteristics of advanced expert finding techniques. 
These include interests, expertise, personalized, and pre-training.
We list them in Table~\ref{tab:methods}.
Our proposed method incorporates all the aforementioned key characteristics of expert finding, which could pre-train more comprehensive personalized characteristics from unsupervised data for improving expert finding.

%% file: 3-method.tex
\section{PROBLEM DEFINITION}

We first formal definition of the expert finding, then give important notations and explanations used throughout the remainder of this paper.

In the expert finding task, the main goal is to match a target question, denoted as $q^t$, with the most appropriate expert who is capable of providing a relevant and accurate answer.
Assuming that there is a candidate expert set $C^u = \{c^u_1, \cdots, c^u_M \}$ and a target question $q^t$ to be answered respectively, where $M$ is the number of experts.
In the set $C^u$, the candidate expert $c_i^u$ is associated with historically answered questions, denoted as $Q_i^u$.
The $Q_i^u$ includes $\{q_1, \cdots, q_n \}$, where $n$ is historical question number.
Each question is represented as a question title including a sequence of words.
The vote scores obtained for each previously answered question of the expert can be presented as $V^u_i = \{v_1, \cdots, v_n \}$.
It is important to note that, the answerer who provides the ``accepted answer'' for a given question is considered the expert, i.e., the ground truth.
Each question has only one ``accepted answer''.

The main notations are summarized in Table~\ref{tab:notation}.

\begin{table*}
\centering
\caption{Notations and Explanations.}
\label{tab:notation}
\begin{tabular}{c|c}
\toprule[1pt]
Notations & Definitions \\
\midrule
$C^u$ & Candidate expert set \\
$q^t$ & Target question \\
$c^u_i$ & $i$-th candidate expert \\
$M$ & Number of experts \\
$Q^u_i$ & $i$-th expert historical answered questions \\
$V^u_i$ & $i$-th expert historical obtained vote scores \\
$n$ & Number of expert historical questions \\
$q_n$ & $n$-th historical question of the expert \\
$v_n$ & $n$-th historical vote score of the expert \\
$d$ & Representation vector dimension \\
$h$ & The number of transformer heads \\
$t_{num}$ & The token number of one input line \\
$\mathcal{V}_w$ & The word vocabulary \\
$\mathcal{V}_s$ & The vote score vocabulary \\
\midrule
$\mathcal{E}$ & The model input information \\
$tk$ & The input token \\
$\mathbf{R}_t$ & The input representation matrix \\
$\mathbf{R}_{token}$ & Token embedding matrix \\
$\mathbf{R}_{seg}$ & Segment embedding matrix \\
$\mathbf{R}_{pos}$ & Position embedding matrix  \\
$\mathbf{R}_{v}$ & The expert vote score embedding matrix \\
$\mathbf{R}_{in}$ & The final expert-level embedding matrix \\
$\mathbf{C}^u$ & Expert embedding matrix \\
$\mathbf{c}^u_i$ & $i$-th expert embedding \\
$\mathbf{e}$ & The expert pre-trained representation \\
$\mathbf{Q}$ & Attention query parameter \\
$\mathbf{K}$ & Attention key parameter \\
$\mathbf{V}$ & Attention value parameter \\
$E_c$ & The matching score predicted by the model \\
$\hat{E}_c$ & Expert-question matching ground truth \\
\midrule
$\mathcal{L}_{ql}$ & Question-level interest loss \\
$\mathcal{L}_{vs}$ & Vote-oriented loss \\
$\mathcal{L}_{wm}$ & Masked language model loss \\
\bottomrule[1pt]
\end{tabular}
\end{table*}

\section{PROPOSED METHOD}

\begin{figure*}
\centering
\includegraphics[width=0.9\textwidth]{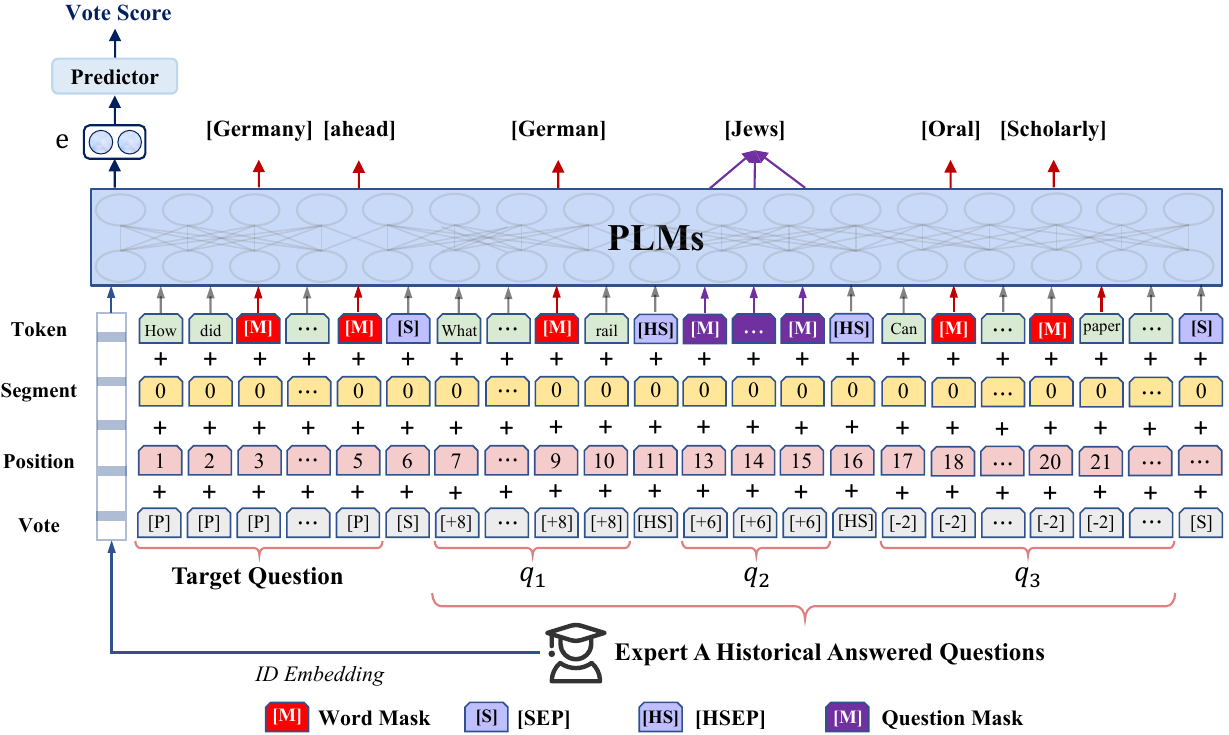}
\caption{\textbf{PEPT Personalized Pre-training Framework}. We unify heterogeneous information into the pre-training model to provide rich information for learning expert representations.
During pre-training, we additionally design question-level masked language model and vote-oriented tasks to empower the model abilities to capture expert interest and expertise.}
\label{fig:framework}
\end{figure*}

In this part, we introduce our method \emph{PEPT}.
Our method including two stages: pre-training expert representations and fine-tuning with the expert finding task. 
These stages are illustrated in the Fig~\ref{fig:framework} and Fig~\ref{fig:ft}.

In the pre-training, we incorporate questions answered by one expert historically and the corresponding target question for constructing target-aware expert-level input units.
Besides, we introduce the expert's personalized information to guide personalized expert pre-training, which could help capture the expert's unique characteristics.
For pre-training, we design the question-level MLM and the vote-oriented tasks for leading the question-level expert interest and expertise learning.
The question-level MLM task is employed to forecast the masked question by taking into account the contextual questions, which could better understand the historical question relatedness for modeling interests.
The vote-oriented task integrates vote scores from the corresponding historical questions (excluding the target question) and enables the model to predict the expert's likely vote score for one target question.

Beneficial from our designed pre-training paradigm, the inputs to the model are very similar between pre-training stage and downstream fine-tuning stage.
And hence, we can directly employ the pre-trained model for conducting expert finding task without any model architecture modification.
The subsequent subsections describe each phase in detail.

\subsection{Pre-training Stage}

The main difference between this work and the conference version is the design of the pre-training architecture.
In this work, we introduce personalized expert information into the model through the soft prompt to achieve personalized pre-training.
Via adopting this approach, the model can be capable of capturing and the distinct characteristics of each expert.
Additionally, by designing specific question-level pre-training tasks, the model gains the ability to capture more fine-grained expert interests and capabilities.

In the following sections, we introduce the PEPT from three perspectives: (1) Input layer, for introducing the input information constructed for the model; (2) Model architecture, for introducing the model components; (3) Pre-training task, for introducing our specially tailored training tasks for expert finding.

\subsubsection{Input Layer}

Different from traditional language knowledge pre-training in Nature Language Processing, which concentrates on corpus-level information input, we aim to empower the BERT abilities to personalized capture expert representations.
We introduce the input layer from the three separate aspects (i.e., interest, expertise, and personalized).

\paragraph{Interest Information} In CQA websites, the user's interests can be reflected from the questions he/she has answered historically.
And hence, as shown in Fig~\ref{fig:framework}, we construct a unified expert-level sequence by directly concatenating the words from the expert's previously answered questions, allowing the model to capture the expert's interest information.
Besides, experts’ representations should be different when faced with different target questions.
Therefore, we further incorporate the target question with the expert-level interest information.

For prompting our model to clearly identify the target question and each previously answered question during pre-training, we design two special tokens and add them to the input sequence.
\begin{itemize}
    \item \texttt{[HSEP]}. This special token is designed to separate different historical questions from one expert.
    \item \texttt{[SEP]}~\cite{devlin2018bert}. We utilize that to distinguish target questions and expert's historical questions.
\end{itemize}
And hence, the input information constructed for the $i$-th expert can be formulated as:
\begin{equation}
\begin{split}
    \mathcal{E} = [\underbrace{[{tk}_1], [{tk}_2], [{tk}_3]}_{q^t}, \mathtt{[SEP]}, \\
    \underbrace{\underbrace{[{tk}_1], [{tk}_2]}_{q_1}, \mathtt{[HSEP]}, \underbrace{[tk_1], [tk_2], [tk_3]}_{q_2}}_{c^u_i}, \mathtt{[SEP]}]  
\end{split}
\end{equation}

Subsequently, taking into account the extensive linguistic knowledge already captured by the original pre-trained BERT weight~\cite{devlin2018bert}, we employ the bert-base-uncased to initialize the model parameters.
And the input representation matrix $\mathbf{R}_t \in \mathcal{R}^{t_{num} \times d}$ ($t_{num}$ is the number of tokens) is generated by summing its corresponding embeddings:
\begin{equation}
    \mathbf{R}_t = \mathbf{R}_{token} + \mathbf{R}_{seg} + \mathbf{R}_{pos} \ ,
\end{equation}
where $\mathbf{R}_{token}$ is the token embedding, and $\mathbf{R}_{seg}$ is the segment embedding, and $\mathbf{R}_{pos}$ is the position embedding.

\paragraph{Expertise Information} In general, the vote scores received by an expert can indicate their expertise in answering different questions. 
A higher vote score reflects a stronger professional expertise in addressing such questions.
To effectively measure expert abilities in different question fields, we incorporate vote scores into our pre-training process.

Specifically, we pre-process vote scores (introduced in Implementation Details) of questions and employ an embedding matrix to map each vote score as an embedding vector (with $d$ as dimension).
To accurately capture the expert's expertise in a specific question, we ensure that the dimension of each vote score embedding matches the dimensionality of the corresponding historical question.
In other words, the length of one question token is $x$, and the dimension of corresponding vote score embedding matrix is $x \times d$.
Then we stack these vote score embeddings as a matrix $\mathbf{R}_v \in \mathcal{R}^{t_{num} \times d}$.
Afterwards, we integrate $\mathbf{R}_v$ with the input representation $\mathbf{R}_t$ as follows:
\begin{equation}
    \mathbf{R}_{in} = \mathbf{R}_{t} + \mathbf{R}_{v} \ .
\end{equation}

It is noted that we employ the special token \texttt{[PAD]} to replace the corresponding vote score of the target question for prediction.

\paragraph{Personalized Information} Different from NLP, expert finding further emphasize personalization.
User ID embeddings are commonly used in recommendation models and could be considered as the identifying information of individual users~\cite{rendle2009bpr}.
Hence, we introduce the expert ID into the pre-training, which could reflect unique expert characteristics and help personalized pre-training.
A simple and intuitive approach is to consider expert IDs as distinct word tokens and incorporate them into the pre-trained model's vocabulary.
However, considering potentially millions of experts on the CQA websites, this would lead to an excessively large word vocabulary, making pre-training time-consuming.
It's worth noting that the expert ID doesn't need to be in word form and can be represented by a vector embedding created through random initialization or other techniques.
Therefore, we treat expert IDs as additional prompting tokens rather than adding them to the vocabulary.

Specifically, we prepare one token embedding matrix
$\mathbf{C}^u \in \mathcal{R}^{M \times d}$, where $M$ is the number of experts.
Then, $i$-th expert's vector representation is retrieved by:
\begin{equation}
    \mathbf{c}^u_i = \mathbf{C}^u f(i) , \
\end{equation}
where $\mathbf{c}^u_i \in \mathcal{R}^{1 \times d}$, and $f(i) \in {\{1,0\}}^M$ represents a one-hot vector, where the $1$ indicates the position of the expert's vector in $\mathbf{C}^u$.
Then, we concatenate the ID embedding of the expert at the begin of the corresponding $\mathbf{R}_{in}$, which is denoted as $[\mathbf{c}^u_i;\mathbf{R}_{in}]$ with $\mathcal{R}^{(t_{num}+1) \times d}$ as dimension.
This could prompt the model to personalized pre-train expert interest and expertise across various questions.

\subsubsection{Model Architecture}

BERT~\cite{devlin2018bert} contains multiple bidirectional Transformer encoders. 
Within each layer, self-attention and feed-forward operations are conducted to capture contextual information and generate contextualized representations of the input tokens.

\paragraph{Self-Attention Layer} Self-attention plays a vital role as a fundamental building block within the Transformer architecture. 
By enabling each token to attend to all other tokens in the sequence, BERT effectively captures contextual relationships between the tokens. The self-attention function is described as follows:
\begin{equation}
    Att(\mathbf{Q}, \mathbf{K}, \mathbf{V}) = Softmax(\mathbf{Q}\mathbf{K}^T/\sqrt{d})\mathbf{V} \ .
\end{equation}

Besides, the BERT employs multi-head attention, where self-attention is performed in parallel multiple times. 
This function \textit{MH}($\cdot$) allows the model to capture different types of contextual relationships, which can be denoted as:
\begin{equation}
    MH(\mathbf{R}_{in}) = [{head}_1;\cdots;{head}_h]\mathbf{W} \ ,
\end{equation}
\begin{equation}
    {head}_i = Att(\mathbf{R}_{in}\mathbf{W}^q_i, \mathbf{R}_{in}\mathbf{W}^k_i, \mathbf{R}_{in}\mathbf{W}^v_i) \ ,
\end{equation}
where $\mathbf{W}^q_i, \mathbf{W}^k_i, \mathbf{W}^v_i \in \mathcal{R}^{d \times \frac{d}{h}}$ and $\mathbf{W} \in \mathcal{R}^{d \times d}$ are learnable parameters.
Through this function, the output is $\mathbf{H} \in \mathcal{R}^{(t_{num}+1) \times d}$ ($t_{num}$ is the token number), which capture the contextual relationships.

\paragraph{Feed-Forward Layer} Then, we utilize feed-forward layers within each transformer encoder layer to apply non-linear transformations to the representations captured by the self-attention mechanism.
Given the input matrix $\mathbf{H}$, the calculation of this function is defined as:
\begin{equation}
    FFN(\mathbf{H}) = RELU(\mathbf{H}\mathbf{W}^f_1 + b^f_1)\mathbf{W}^f_2 + b^f_2 \ ,
\end{equation}
where $\mathbf{W}^f_1$, $\mathbf{W}^f_2$ and $b^f_1$, $b^f_2$ are learnable parameters.
More details please refer to~\cite{vaswani2017attention}.

\begin{figure*}
\includegraphics[width=0.6\textwidth]{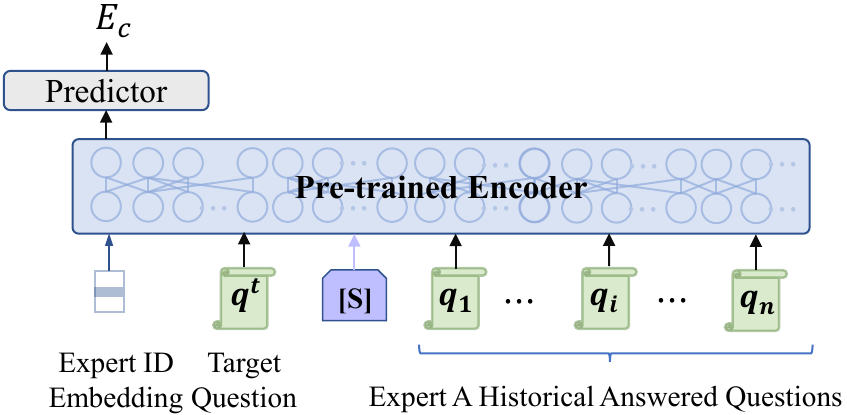}
\caption{\textbf{PEPT Personalized Fine-tuning}. The pre-trained encoder is the pre-trained expert weight from PEPT Personalized Pre-training.}
\label{fig:ft}
\end{figure*}

\subsubsection{Pre-training Task}

We introduce our designed pre-training tasks in this part, which are the core mechanisms that encourage PLMs to model experts' personalized interest and expertise.

\paragraph{Interest Pre-training} In fact, the interest of experts could be reflected in the model input sequence that we design.
As illustrated in Section 1, modeling the question-level semantic information is important for expert interest modeling.
However, existing preliminary works could only capture interests from word-level information rather than question-level semantic information.
As expert histories are diverse and harder to predict, different from the MLM task, we propose an additional pre-training task called question-level masked language model to capture question-level interests.

Specifically, given the example illustrated in Fig~\ref{fig:framework}, inspired by SpanBERT~\cite{hiCLWZL20}, we mask all title words of the question $q_2$,  then predict the token \texttt{[Jews]} in the $q_2$ based on the contextual semantic information.
And the loss function is:
\begin{equation}
    \mathcal{L}_{ql} = - logp(\texttt{Jews}|x_{11}, x_{16}, p_{14}), \texttt{Jews} \in [1,2,\cdots, \lvert\mathcal{V}_w\rvert] \ ,
\end{equation}
where $x_{11}$ and $x_{16}$ represent the contextual historical question semantic information, and $p_{14}$ represents the position of the token that is predicted by the model.

\paragraph{Expertise Pre-training} As denoted above, the vote score the expert received could reflect the expertise in answering one question.
Generally, higher vote scores indicate higher expertise, and the answer would be more satisfied the CQA users.
Therefore, we design a vote-oriented task to pre-train the model and enhance its capacity to discern the expert's expertise in addressing the target question.
Specifically, given the instance shown in Fig~\ref{fig:framework}, the model output embedding from the expert ID can be viewed as an expert-level representation $\mathbf{e} \in \mathcal{R}^{1 \times d}$.
Hence, given a target question, we directly employ $\mathbf{e}$ to predict the vote score that the expert would receive, and the loss function is:
\begin{equation}
    \mathcal{L}_{vs} = - \sum_{i=1}^{\mathcal{S}}logp(v=v_i|\theta), v_i \in [1,2,\cdots, \lvert\mathcal{V}_s\rvert] ,
\end{equation}
where $\mathcal{S}$ is the masked target question score, and $\mathcal{V}_s$ is the vote score vocabulary.

Additionally, since the potent capacity of the original MLM task~\cite{devlin2018bert} to train PLMs, we incorporate this task to acquire linguistic knowledge in CQA websites.
The MLM task randomly masks certain words and subsequently leverages bidirectional contextual information to reconstruct the input sequence.
The formal definition is:
\begin{equation}
    \mathcal{L}_{wm} = - \sum_{i=1}^{\mathcal{N}}logp(n=n_i|\theta), n_i \in [1,2,\cdots, \lvert\mathcal{V}_w\rvert] ,
\end{equation}
where $\mathcal{N}$ is the masked token number, and $\mathcal{V}_w$ is the word vocabulary.

\subsubsection{Model Learning}
During pre-training, the PEPT model is trained with three tasks, and the loss function used is:
\begin{equation}
    \mathcal{L}_{pt} = \mathcal{L}_{ql} + \mathcal{L}_{vs} + \mathcal{L}_{wm} \ .
\end{equation}

Through the integration of these loss functions, our PEPT model effectively pre-trains personalized expert-level representations that encompass not only the interests but also the expertise of the experts.

\subsection{Fine-tuning Stage}

In this section, we perform the expert finding task using a fine-tuning approach.
As depicted in Fig~\ref{fig:ft}, we combine the previously answered questions by the expert with the target question and input the sequence directly into the pre-trained model, which is denoted as $[\mathbf{c}^u_i;\mathbf{R}_{in}]$.
By adopting this approach, we ensure that the inputs during the fine-tuning stage remain consistent with those used during the pre-training phase. 
This consistency helps minimize the disparity between the pre-training and fine-tuning stages.

Next, we add a linear layer to the special embedding $\mathbf{e}$ (derived from the ID embedding) to predict the expert-question matching probability $E_c$, denoted as:
\begin{equation}
    E_c = \mathbf{W}_c(\mathbf{e}^T) + \mu_c \ ,
\end{equation}
where $\mathbf{W}_c$ and $\mu_c$ are model parameters.

During fine-tuning, we follow previous works~\cite{li2019personalized,peng2022towards} and utilize the negative sampling technology~\cite{huang2013learning} to generate $k$ negative samples.
Then, we employ the cross-entropy loss for optimizing, which can be denoted as:
\begin{equation}
E_c = \frac{\exp(E_c)}{\sum_{j=1}^{k+1}\exp(E_j)} \ , Loss = -\sum_{c=1}^{k+1}\hat{E}_clog(E_c) \ ,
\end{equation}
where $\hat{E}_c$ is the ground truth and $E_c$ is the normalized probability predicted by methods.

%% file: 4-experiments.tex
\section{EXPERIMENTAL SETTINGS}

\begin{table*}
\centering
\caption{Statistical Details of Each Dataset.}
\resizebox{0.8\textwidth}{!}{
\begin{tabular}{c|c|c|c|c|c}
\toprule[1pt]
Datasets & \# density (\%) & \# questions & \# answerers & \# answers & \# avg.title length \\
\midrule
Es & 0.0429 & 36,271 & 3,260 & 50,801 & 10.05 \\
Gis & 0.0435 & 50,718 & 3,168 & 70,034 & 9.75  \\
Biology & 0.2081 & 8,704 & 630 & 11,411 & 9.84 \\
English & 0.0468 & 46,692 & 4,781 & 104,453 & 9.68  \\
Electronics & 0.0585 & 56,614 & 3,084 & 102,214 & 9.28 \\
CodeReview & 0.0696 & 36,947 & 2,242 & 57,622 & 7.27  \\
\bottomrule[1pt]
\end{tabular}
}
\label{tab:dataset}
\end{table*}

To evaluate the performance of PEPT, we conduct experiments on six CQA datasets.
First, we introduce the datasets employed in our paper.
Afterwards, we introduce the implementation details of our experiment.
Then, we introduce the baseline methods and evaluation metrics used in this paper.

\subsection{Datasets}

In experiments, we use a dataset comprising 326781 target-aware expert input information extracted from StackExchange\footnote{https://archive.org/details/stackexchange}. 
Each expert corresponds to his/her historical answered questions. 
This is used for pre-training expert representations in PEPT.
We evaluate the effectiveness of the pre-trained model in six domains: \textit{English}, \textit{Biology}, \textit{Es}, \textit{Electronics}, \textit{Gis}, and \textit{CodeReview}.
Each question is accompanied by a sequence of title words that provide question semantic information.
Additionally, each question has an ``accepted answer'' chosen from multiple answers provided by different experts.
The negative sample in training set is set to 9.
In evaluation, the expert who has provided the ``accepted answer'' is the ground truth.
We follow the pre-processing method employed in previous research~\cite{peng2022towards}.
The candidate expert set for question in evaluation includes the original answerers who previously answered the question, and select additional answerers from the pool of answerers (one set contains 20 experts in total).
It is noted that the number of negative samples are same in different methods for fair comparison.
Table~\ref{tab:dataset} provides an overview of the statistical properties of the datasets, offering a summary of their key characteristics.
We split each dataset into three distinct subsets in chronological order: training, validation and testing sets with 80\%, 10\% and 10\% of the dataset, respectively.
Noted that the validation set and the testing set do not appear during pre-training.

\subsection{Implementation Details}

In this section, we introduce the experimental settings and hyperparameters employed in our model.
Next, we introduce the vote score pre-processing.
It is noted that all hyperparameters in our model were fine-tuned using the validation dataset.

\paragraph{Experimental Setting}
Our personalized pre-training (PEPT) model is based on the bert-base-uncased~\cite{devlin2018bert}.
The batch size for both pre-training and fine-tuning is $16$.
For alleviating over-fitting, we use dropout technology~\cite{srivastava2014dropout}. Specifically, the dropout ratio is $0.2$ in pre-training and $0.3$ in fine-tuning.
We optimize our model using the Adam optimization strategy~\cite{kingma2015adam}. 
We set the learning rate to $5e-5$ during pre-training and $5e-6$ during fine-tuning.
The token embedding and expert representation dimensions are both $768$.
The maximum training step during pre-training is 50,000.
We conduct each experiment independently and repeat it three times, and report the average results.
All experiments are carried out using the PyTorch framework.
We utilize a single GPU server with an Intel(R) Xeon(R) CPU@2.20GHz and two RTX-3090 GPUs, each with 24GB of memory.

\paragraph{Pre-processing of Vote scores}
In this part, we introduce how to process the vote scores of the questions.
One of the key challenges we encountered is a large variance in the distribution of vote scores.
For example, in the Biology dataset, we observe the maximum obtained vote score of one question is $+287$, while the minimum is $-8$, and the average vote score is $3.49$.

Therefore, to alleviate this, we employ a normalization approach on a per-question basis.
First, we translate the all vote scores to eliminate negative values (i.e., all vote scores are added to the minimum value).
Subsequently, we apply a logarithmic transformation (i.e., $ln(\cdot)$) to the vote scores of questions to reduce the impact of excessive variance.
This helps ensure that the obtained scores of questions are more evenly distributed across the dataset.
To simplify data processing and ensure roughly equivalent numbers of scores within each score segment, we normalized the vote scores to integers ranging from $0$ to $9$.
The normalization process is conducted as follows:
\begin{equation}
    v_{min} = min(V), v_{max} = max(V) \ ,
\end{equation}
where $v_{min}$ and $v_{max}$ represent the minimum score and the maximum score of the $V$ (all question vote scores) respectively.
\begin{equation}
    v^*_i = round(\frac{9}{v_{max}-v_{min}} * (v_i - v_{min}) + 1) \ ,
\end{equation}
here, $v^*_i$ represents the normalized score of the $v_i$ in $V$, and $round$ denotes the rounding operator.

\subsection{Comparison Methods}

We compare the PEPT with two different types of baselines:

\textbf{First}, the neural expert finding methods that utilize various neural networks to learn expert and question representations, then use these representations for finding experts.
\begin{itemize}
    \item \textbf{CNTN}~\cite{qiu2015convolutional} employs the CNN to model question semantic information and model experts, then calculate the question-expert relevance scores.
    \item \textbf{NeRank}~\cite{li2019personalized} learns question, raiser and expert representations via a Heterogeneous Information Network (HIN)~\cite{sun2011pathsim} algorithm, then uses the CNN network to find the expert for questions.
    \item \textbf{TCQR}~\cite{zhang2020temporal} incorporates a temporal context-aware model that captures temporal patterns in experts' answering behavior across multiple time scales.
    \item \textbf{RMRN}~\cite{fu2020recurrent} incorporated reasoning memory cells (RMCs) to capture the essence of the question and leveraged an attention mechanism to extract relevant information from the historical records of candidate experts
    \item \textbf{UserEmb}~\cite{ghasemi2021user} utilizes a node2vec~\cite{grover2016node2vec} for learning structural relationships among users and uses a word2vec~\cite{mikolov2013distributed} to model the semantic context of words, and integrates the social features and word features to improve the expert finding.
    \item \textbf{EFHM}~\cite{peng2022towardskbs} is a multi-grained (word-level, question-level and expert-level) hierarchical matching framework for expert finding in CQA websites, which could capture more fine-grained matching relationships and achieve superior performance.
    \item \textbf{PMEF}~\cite{peng2022towards} designs a multi-view attentive expert finding method with a personalized mechanism, which could comprehensively model expert/question and precisely match them.
\end{itemize}

\textbf{Second}, the latest PLM-based methods model expert or question features using pre-trained language models.
\begin{itemize}
    \item \textbf{ExpertBert}~\cite{liu2022expertbert} is a preliminary pre-training framework for expert finding. The framework is designed to accurately model the question-expert matching, which could improve performance of expert finding.
    \item \textbf{ExpertPLM}~\cite{peng2022expertplm} designs an expert-specific pre-training paradigm for finding expert. The key feature of ExpertPLM is a reputation-augmented masked language model that is designed to model an experts' interests and expertise based on their previously answered questions and reputations.
\end{itemize}

Some papers have already published their source code on Github, such as NeRank~\cite{li2019personalized}, PMEF~\cite{peng2022towards}, etc. 
For these models, we download the code and check the model details and parameters based on their original paper to ensure that the model is consistent with the original paper.
For models without publicly available source code, such as RMRN~\cite{fu2020recurrent}, TCQR~\cite{zhang2020temporal}, etc. 
We follow technical details reported by the paper to reproduce the model and set model hyperparameters based on the hyperparameter settings reported in the paper.

\subsection{Evaluation Metrics}

For evaluating the model performance, we use the following three widely employed evaluation metrics in recommendation:

\begin{itemize}
    \item Mean Reciprocal Rank (MRR): Measures models in identifying the most appropriate relevant expert from a candidate expert set. It is computed as the reciprocal of the rank assigned to the first relevant expert, which is denoted as:
    \begin{equation}
        {\rm MRR} = \frac{1}{N}\sum_{i=1}^{N}\frac{1}{r_i} \ ,
    \end{equation}
    where $N$ represents the total number of target questions and $r_i$ refers to the rank assigned to the most suitable expert within the candidate expert set for a specific question.
    \item Precision@$K$: We measure the proportion of target questions where the true expert is ranked within the top-$K$ positions among the candidate answerers. In our experiments, we set  $K$ as 1 and 3.
    \item Normalized Discounted Cumulative Gain ($NDCG$@$K$)~\cite{jarvelin2002cumulated}: The metric takes into account both the relevance of the recommended experts and their respective positions within the candidate expert set. It is calculated by summing the discounted relevance scores of the recommended items and normalizing it by the ideal $DCG$ score. And the $NDCG$ is computed as: $NDCG$@$K$ $= \frac{{DCG}_K}{{IDCG}_K}$. 
    The ${DCG}_K$ is computed as:
    \begin{equation}
        {DCG}_K = \sum_{i=1}^K\frac{r_i}{log_2(i+1)} \ ,
    \end{equation}
    where $r_i = 1$ when the $i$-th expert is the ground truth.
    ${IDCG}_K$ represents the ideal discounted cumulative gain.
    And then $NDCG$@$K$ is computed as: $NDCG@K = \frac{{DCG}_K}{{IDCG}_K}$.
    We set $K = 20$.
\end{itemize}

Overall, the above metrics are used to assess the ranking accuracy of the expert recommended by each model.
A higher value for each of these metrics indicates better performance of the model in accurately predicting the most relevant expert.

%% file: 5-results.tex
\section{EXPERIMENTAL RESULTS}

In this section, we represent the experiments and conduct a detailed analysis of PEPT.
Specifically, 
(1) Performance Comparison compares the PEPT with baselines;
(2) Ablation Study explores how different parts of the model affect model performance;
(3) Pre-trained Expert Representation Analysis analyzes how the pre-trained expert representations benefit the expert finding.
(4) Hyper-parameter Analysis analyzes the model performance with different model hyper-parameters.
(5) Case Study understands PEPT’s advantages in finding experts more intuitively.

\subsection{Performance Comparison}

\begin{table*}
\centering
\caption{Expert finding performance ($\%$) of different models. Among the baselines, the underlined results indicate the best performance.
We conduct t-tests and show that our proposed method outperforms the other baselines with a significant p-value $<$ 0.05.}
\resizebox{0.85\textwidth}{!}{
\begin{tabular}{c|c|c|c|c|c|c|c|c|c|c|c|c}
\toprule[1.5pt]
Dateset & \multicolumn{4}{c|}{\textbf{English}} & \multicolumn{4}{c|}{\textbf{Gis}} & \multicolumn{4}{c}{\textbf{CodeReview}} \\
\midrule
\diagbox{Method}{Metric} & MRR & P$@$1 & P$@$3 & NDCG$@$20 & MRR & P$@$1 & P$@$3 & NDCG$@$20 & MRR & P$@$1 & P$@$3 & NDCG$@$20 \\
\midrule
CNTN & 29.68 & 18.37 & 39.36 & 42.25 & 40.12 & 23.12 & 40.15 & 42.01 & 35.53 & 20.13 & 50.77 & 50.35 \\
NeRank & 48.95 & 27.16 & 61.43 & 56.41 & 46.97 & 30.32 & 5577 & 58.36 & 49.47 & 30.89 & 60.55 & 61.54 \\
TCQR & 34.25 & 19.27 & 49.87 & 48.22 & 45.53 & 26.34 & 5489 & 56.37 & 42.53 & 21.12 & 53.82 & 58.39 \\
RMRN & 46.77 & 25.22 & 61.62 & 56.75 & 48.97 & 32.39 & 57.77 & 58.32 & 43.11 & 25.17 & 55.80 & 58.92 \\
UserEmb & 31.73 & 19.56 & 42.36 & 45.51 & 32.23 & 24.33 & 44.77 & 45.62 & 39.15 & 20.31 & 51.26 & 52.46 \\
EFHM & 50.87 & 31.66 & 62.16 & 57.65 & 47.28 & 32.31 & 57.68 & 59.05 & 49.22 & 30.25 & 60.58 & 59.77 \\
PMEF & 49.47 & 28.65 & 63.14 & 58.75 & 48.61 & 33.11 & 58.88 & 59.11 & 50.20 & 31.39 & 61.61 & 61.70 \\
\midrule
ExpertBert & 52.01 & \underline{32.32} & 64.97 & \underline{62.38} & \underline{51.70} & \underline{35.37} & \underline{60.66} & \underline{62.65} & \underline{51.23} & \underline{32.87} & 62.65 & \underline{62.48} \\
ExpertPLM & \underline{52.50} & 31.68 & \underline{65.12} & 61.17 & 50.85 & 34.56 & 60.58 & 61.78 & 51.15 & 32.63 & \underline{63.13} & 62.42 \\
\midrule
\textbf{PEPT} & \textbf{53.22} & \textbf{33.28} & \textbf{65.66} & \textbf{63.18} & \textbf{52.81} & \textbf{35.99} & \textbf{61.16} & \textbf{62.95} & \textbf{52.88} & \textbf{35.27} & \textbf{63.65} & \textbf{63.58} \\
\midrule[1.5pt]
Dateset & \multicolumn{4}{c|}{\textbf{Es}} & \multicolumn{4}{c|}{\textbf{Biology}} & \multicolumn{4}{c}{\textbf{Electronics}} \\
\midrule
\diagbox{Method}{Metric} & MRR & P$@$1 & P$@$3 & NDCG$@$20 & MRR & P$@$1 & P$@$3 & NDCG$@$20 & MRR & P$@$1 & P$@$3 & NDCG$@$20 \\
\midrule
CNTN & 38.97 & 27.20 & 40.78 & 46.93 & 33.58 & 20.17 & 33.69 & 36.91 & 49.52 & 27.40 & 65.38 & 61.46 \\
NeRank & 56.47 & 39.32 & 64.13 & 66.17 & 43.71 & 28.86 & 53.61 & 49.91 & 51.05 & 34.59 & 59.76 & 62.21 \\
TCQR & 47.16 & 30.59 & 59.41 & 60.53 & 39.62 & 24.06 & 4422 & 47.16 & 50.16 & 32.33 & 59.53 & 60.04 \\
RMRN & 55.16 & 37.61 & 63.04 & 64.53 & 46.62 & 31.53 & 55.62 & 55.78 & 56.22 & 40.38 & 67.59 & 67.29 \\
UserEmb & 47.03 & 30.32 & 57.35 & 48.69 & 34.97 & 22.63 & 36.75 & 46.72 & 41.75 & 29.54 & 45.41 & 51.70 \\
EFHM & 55.28 & 40.61 & 63.17 & 65.03 & 48.50 & 29.23 & 57.66 & 57.58 & 57.83 & 41.36 & 68.50 & 67.38 \\
PMEF & 57.35 & 42.69 & 65.20 & 66.61 & 47.19 & 32.16 & 56.93 & 56.88 & 58.72 & 42.38 & 69.50 & 68.29 \\
\midrule
ExpertBert & 57.99 & 42.78 & 65.37 & 66.81 & \underline{49.70} & 33.23 & 57.66 & 58.65 & 59.23 & 42.57 & 69.68 & 69.33 \\
ExpertPLM & \underline{58.40} & \underline{43.80} & \underline{65.88} & \underline{67.93} & 49.56 & \underline{33.53} & \underline{58.68} & \underline{59.29} & \underline{60.03} & \underline{43.56} & \underline{70.25} & \underline{70.46} \\
\midrule
\textbf{PEPT} & \textbf{59.66} & \textbf{44.76} & \textbf{66.97} & \textbf{68.38} & \textbf{51.06} & \textbf{34.53} & \textbf{60.00} & \textbf{62.13} & \textbf{61.23} & \textbf{44.70} & \textbf{71.33} & \textbf{71.98} \\
\bottomrule[1.5pt]
\end{tabular}
}
\label{tab:result}
\end{table*}

Table~\ref{tab:result} shows the overall results of our proposed model in comparison to other baselines.
The experimental results reflect following observations:
\begin{enumerate}
    \item Early methods, such as CNTN, exhibit poor performance on almost all datasets. 
    This attributes to the fact that these methods rely on the max or mean operator on experts' histories to obtain expert features, which fails to capture the different importance of different historical questions in modeling the interest of an expert.
    \item In contrast, recent attention-based methods, such as RMRN and PMEF, have achieved better results. The reason is that these methods leverage attention mechanisms to selectively weigh the importance of different historical questions in modeling expert representations.
    It is worth noting that the PMEF outperforms most of the baselines.
    This is because PMEF utilizes a multi-view attentive paradigm, enabling it to learn more comprehensive representations for experts and questions.
    \item Pre-training based methods (ExpertBert, ExpertPLM, PEPT) always outperform other methods. These methods employ pre-training technology to capture the general expert/question knowledge based on unsupervised data and transfer them to the expert finding.
    For example, ExpertPLM utilized a reputation-augmented MLM to capture the coarse-grained expert characteristics.
    And the model could capture the universal information from unlabeled data in a task-agnostic way, which can help learn better question-expert relatedness in expert finding.
    \item Our proposed method PEPT surpasses two very recent works ExpertBert~\cite{liu2022expertbert} and ExpertPLM~\cite{peng2022expertplm}. The reason is that the ExpertBert could not capture expert abilities to answer different questions and the ExpertPLM could not capture more fine-grained question-level interest and expertise.
    And both of them are incapable of modeling personalized expert representations, which might fail to accurately model different experts.
\end{enumerate}

In summary, the results validate the motivation proposed in Section 1 and reflect that PEPT with all the characteristics in the Table~\ref{tab:methods} can achieve the best performance.

\subsection{Ablation Study}

\begin{table*}
\centering
\caption{Characteristics of Different Variants. The variants possesses the feature marked with a check mark, while the feature marked with a cross is not possessed by the variant.}
\label{tab:variants}
\resizebox{0.8\textwidth}{!}{
\begin{tabular}{c|c|c|c|c|c|c}
\toprule[1pt]
\diagbox{Characteristics}{Variants} & FT Bert & Only Em & w/o Expertise & w/o Interest & w/o Personalized & PEPT \\
\midrule
Expert-level Information & $\surd$ & $\surd$ & $\surd$ & $\surd$ & $\surd$ & $\surd$ \\
Vote Score & $\times$ & $\times$ & $\times$ & $\surd$ & $\surd$ & $\surd$ \\
ID Embedding & $\times$ & $\times$ & $\surd$ & $\surd$ & $\times$ & $\surd$ \\
\midrule
MLM & $\times$ & $\surd$ & $\surd$ & $\surd$ & $\surd$ & $\surd$ \\
Vote-oriented Task & $\times$ & $\times$ & $\times$ & $\surd$ & $\surd$ & $\surd$ \\
Question-level MLM & $\times$ & $\times$ & $\surd$ & $\times$ & $\surd$ & $\surd$ \\
\bottomrule[1pt]
\end{tabular}
}
\end{table*}

To further explore how different modules of the pre-training model impact the model performance, we design five variants for experimenting (CodeReview and Gis datasets as examples), including:
\begin{enumerate}
    \item \textbf{FT Bert}, we directly fine-tune the original BERT~\cite{devlin2018bert} on the CQA expert finding task, without further pre-training.
    \item \textbf{Only Em}, we adopt the original MLM task~\cite{devlin2018bert} to further pre-train the bert-base-uncased model over the CQA expert-level corpus (i.e., one expert histories one input line);
    \item \textbf{w/o Expertise}, we incorporate expert interest modeling and personalized information during the pre-training. Excluding the vote score information and vote-oriented task from our model; 
    \item \textbf{w/o Interest}, we introduce the expert expertise modeling and personalized information during pre-training, but remove the question-level masked language modeling task;
    \item \textbf{w/o Personalized}, we incorporate the expert interest and expertise modeling but remove the personalized information during pre-training.
\end{enumerate}

A general comparison of the five variants is provided in Table~\ref{tab:variants}, and the experimental results are illustrated in Fig~\ref{fig:ablation}. 
Based on these results, we have the following observations:

\begin{figure*}
\centering
\includegraphics[width=0.8\textwidth]{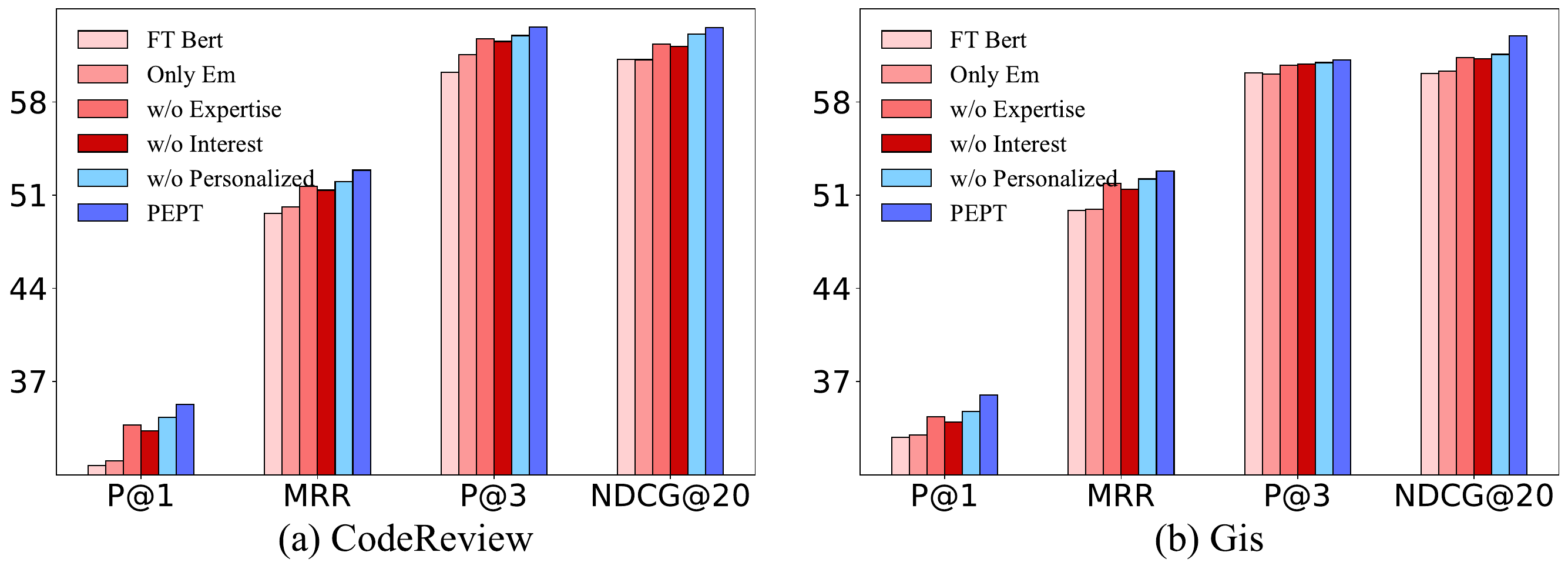}
\caption{Ablation Study. We remove specific components in the PEPT during pre-training to evaluate their different effects on the overall performance. Detailed variants information can be found in Table~\ref{tab:variants}.}
\label{fig:ablation}
\end{figure*}


\begin{enumerate}
    \item Only Em obtains the better performance than FT Bert on the most metrics.
    This is because the Only Em uses expert-level word sequences (i.e., historical answered questions) to further pre-train the model, which could capture more CQA general knowledge for improving model performance.
    \item w/o Expertise and w/o Interest show different degrees of decreased model performance. These results reflect the importance of capturing expert interest and capability for finding suitable experts. Interestingly, the performance of w/o Interest is worse than that of w/o Expertise. We argue that interest may be more important than ability in expert finding. In other words, experts in real life might first consider their interests rather than their abilities before answering questions;
    \item Personalized information is crucial for modeling experts. Removing the personalized information during pre-training leads to a decrease in model performance, which can be seen in w/o Personalized;
    \item Analysis of FT Bert. In fact, directly fine-tuning the BERT for expert finding can only find experts based on text similarity matching. From Table~\ref{tab:variants}, we can find the characteristic difference between the model variants and the FT Bert. From Fig~\ref{fig:ablation}, we can find the performance of model variants is better than the FT Bert.
    Specifically:
    (1) When comparing the model without expertise information (w/o Expertise) and FT BERT, we can find w/o Expertise is superior to FT Bert. This is because the w/o Expertise can effectively capture the experts' personalized fine-grained question-level interest information during pre-training;
    (2) Similarly, the model without interest information (w/o Interest) is also better that FT Bert. This phenomenon reveals that w/o Interest can successfully capture the experts' personalized expertise information for the target question during pre-training;
    (3) Lastly, the comparison between the model without personalized information (w/o Personalized) and FT Bert shows that w/o Personalized is capable of simultaneously capturing both expert expertise and interest information during pre-training.
\end{enumerate}

Overall, the ablation studies meet our motivation and demonstrate the effectiveness of each component in our proposed model.

\subsection{Pre-trained Expert Representation Analysis}

Our pre-training paradigm aims to learn universal and initial expert representations from unsupervised data, which can be transferred to the downstream expert finding task.
We conduct additional experiments aimed at exploring the influence of pre-trained expert representations.

\subsubsection{Effect of Alleviating Data Sparsity}

In real-world applications, expert finding often faces the challenge of data sparsity due to the need for amounts of supervised data.
Our pre-training expert finding paradigm could have the potential to alleviate this problem.
In this section, we investigate whether our pre-training paradigm can alleviate this problem.
To simulate the scenario of data sparsity, we conduct experiments by varying the proportions of training datasets used during the fine-tuning process, namely $[100\%, 80\%, 60\%, 40\%, 20\%]$.
Noted that the validation data and testing data used in these experiments are the same as those used in the main experiments.
We select the PEPT, ExpertPLM, and PMEF for comparison.

\begin{figure*}
\centering
\includegraphics[width=0.8\textwidth]{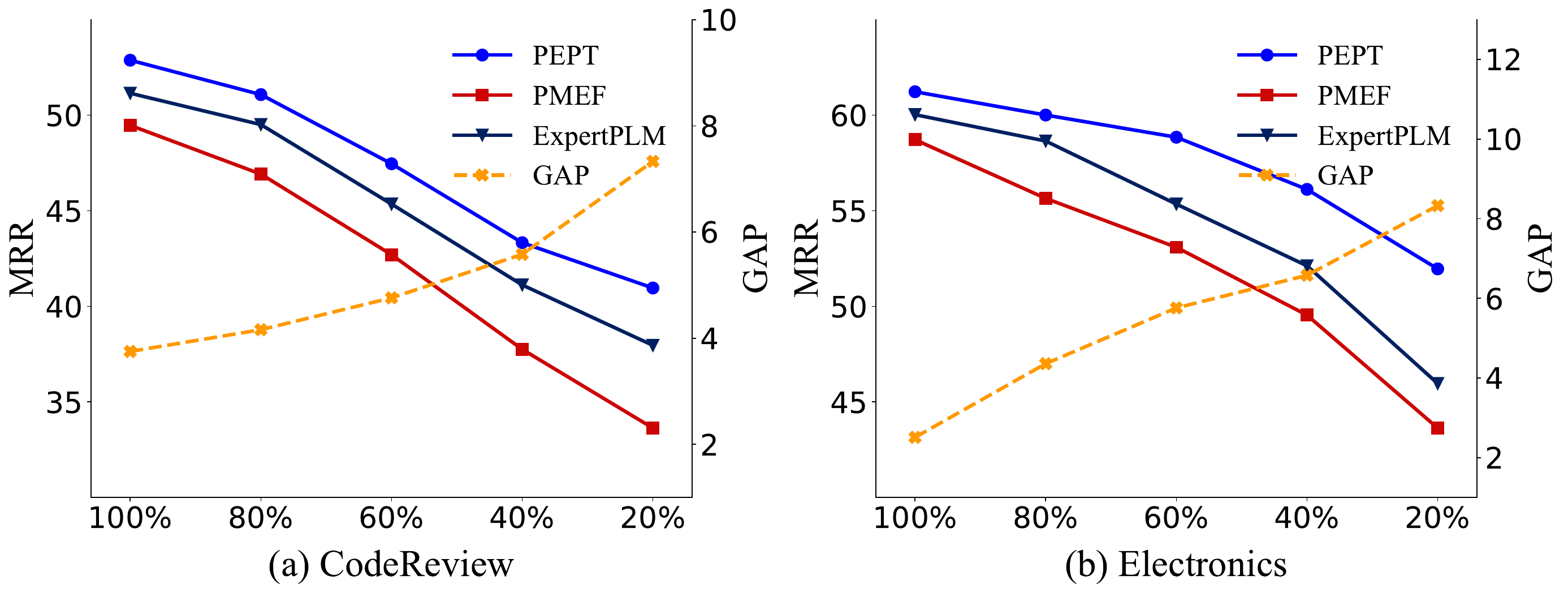}
\caption{Impacts of Fine-tuning Data Ratio. The abscissa is the sparseness of the dataset, and the left ordinate is the MRR metric that measures model performance. GAP = PEPT - PMEF.}
\label{fig:traindata}
\end{figure*}

Fig~\ref{fig:traindata} shows the results on CodeReview and Electronics datasets.
The performance of all models substantially drops when a smaller amount of training data is used.
However, our pre-trained model consistently outperforms the best non pre-trained model PMEF.
Besides, we observe that with decreasing training data, the performance gap between PEPT and PMEF widens.
This reflects that the advantage of our pre-trained representations becomes even more pronounced when the training data is limited.
This is because our pre-training paradigm enables the model to personalized leverage expert behaviors and vote scores for modeling experts' interests and expertise.
Besides, it could reduce the model's dependence on training data.

Furthermore, we can find that although ExpertPLM also has the ability to alleviate data sparsity issues, it is still inferior to PEPT.
This is because ExpertPLM can only pre-train coarse-grained expert representations, and it still needs sufficient supervised data for fine-tuning to capture specific representations of different experts. 
Different from that, PEPT can learn personalized expert representations during pre-training and capture more fine-grained characteristics for expert finding.
Overall, our experimental results validate that PEPT could mitigate the impact of the data sparsity problem in expert finding.


\begin{figure*}
\centering
\includegraphics[width=0.8\textwidth]{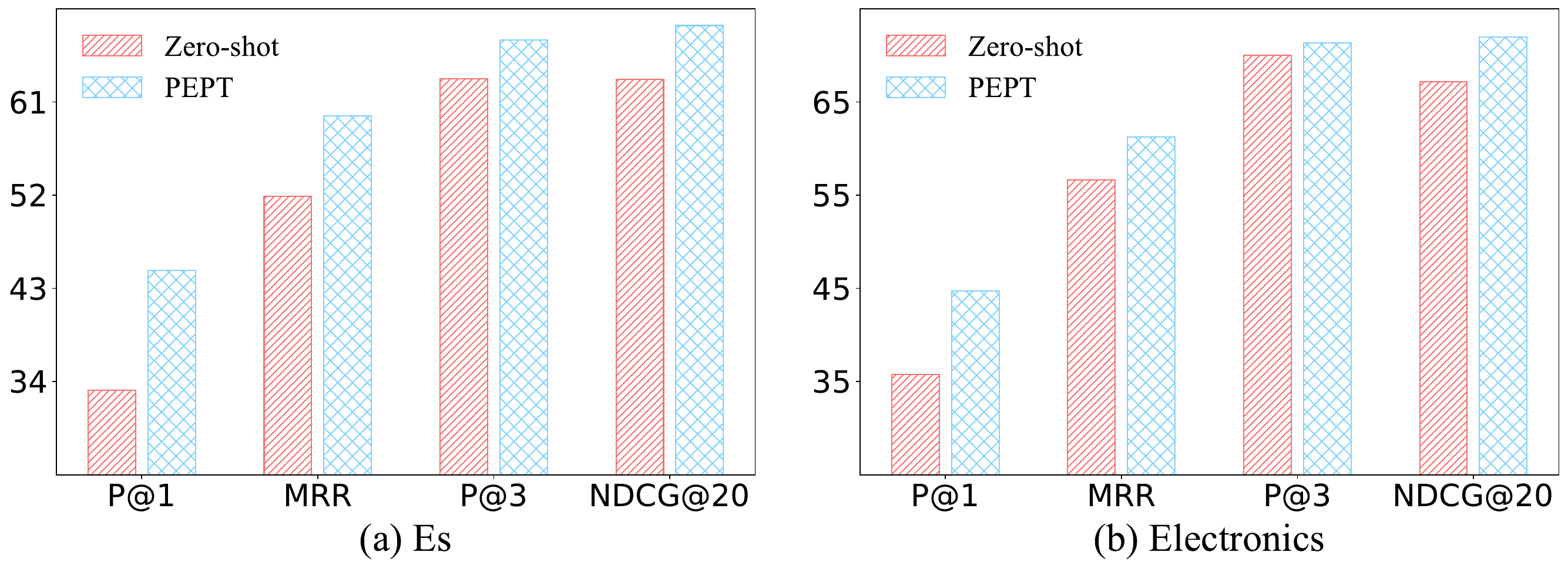}
\caption{Zero-shot Performance.}
\label{fig:zeroshot}
\end{figure*}

\begin{figure*}
\centering
\includegraphics[width=0.85\textwidth]{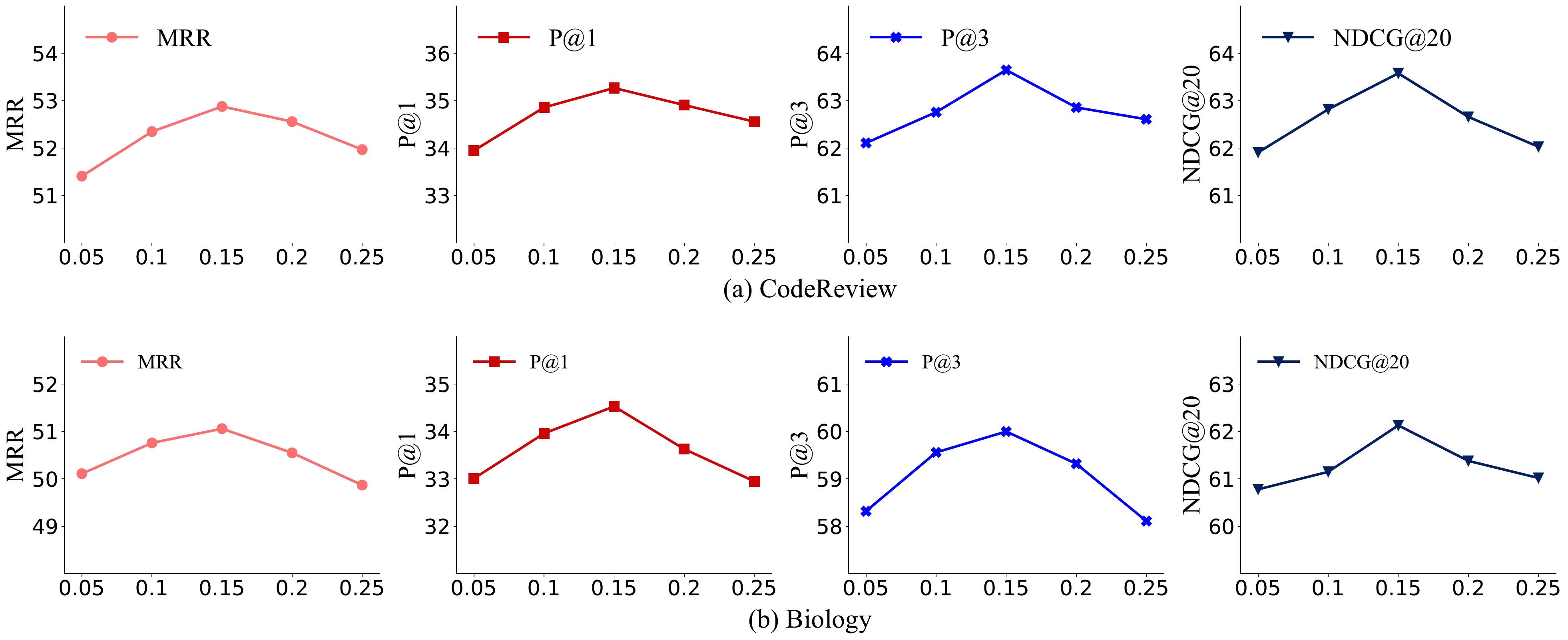}
\caption{Impacts of Different Word Masking Ratios.}
\label{fig:mlmratio}
\end{figure*}




\subsubsection{Zero-shot Performance}

In this part, we test the model performance with the zero-shot setting from Biology to Es and Electronics (we fine-tune the model based on the Biology dataset, and test the performance on the Es and Electronics). 
The experimental results are shown in Fig~\ref{fig:zeroshot}.

It is surprising that the model can still achieve good performance (especially in P@3) by simply transferring from Biology to Es/Electronics without any fine-tuning. The reason could be that the model has already captured a large amount of expert and question related knowledge during pre-training, and merely fine-tuning on smaller datasets (e.g., biology) could stimulate model understanding of the expert finding task, leading to zero-shot performance on other large datasets.

\subsection{Further Analysis}

In this section, we conduct experiments to evaluate the effect of key hyperparameters (including Masked Word Ratio, Masked Question Ratio, Learning Rate in Fine-tuning) in our approach.

\subsubsection{Masked Word Ratio}

The masked word ratio is a hyperparameter in our proposed PEPT model during the expert representation pre-training.
To investigate the sensitivity of that in the pre-training, we vary the ratio from $0.05$ to $0.25$ in increments of $0.05$.

The results are presented in Fig~\ref{fig:mlmratio}. 
All metric results initially improve as the mask ratio increases, reaching the optimal model performance at a certain point, and then gradually declining.
If the word mask ratio is set too small, the pre-training process may not capture enough CQA language knowledge and result in reduced performance in the downstream task.
A high mask ratio during pre-training may lead to excessive \texttt{[MASK]} symbol occurrence, causing a mismatch and negatively impacting performance during fine-tuning.
Hence, during the pre-training stage, a masked word ratio of $0.15$ is selected.

\subsubsection{Masked Question Ratio}

In our method, we design a question-level pre-training task for modeling expert interests.
To explore the impacts of different masking question ratios on model performance, we vary the question masking ration in $[10\%, 15\%, 20\%, 25\%, 30\%]$ for experiments.

\begin{figure*}
\centering
\includegraphics[width=0.8\textwidth]{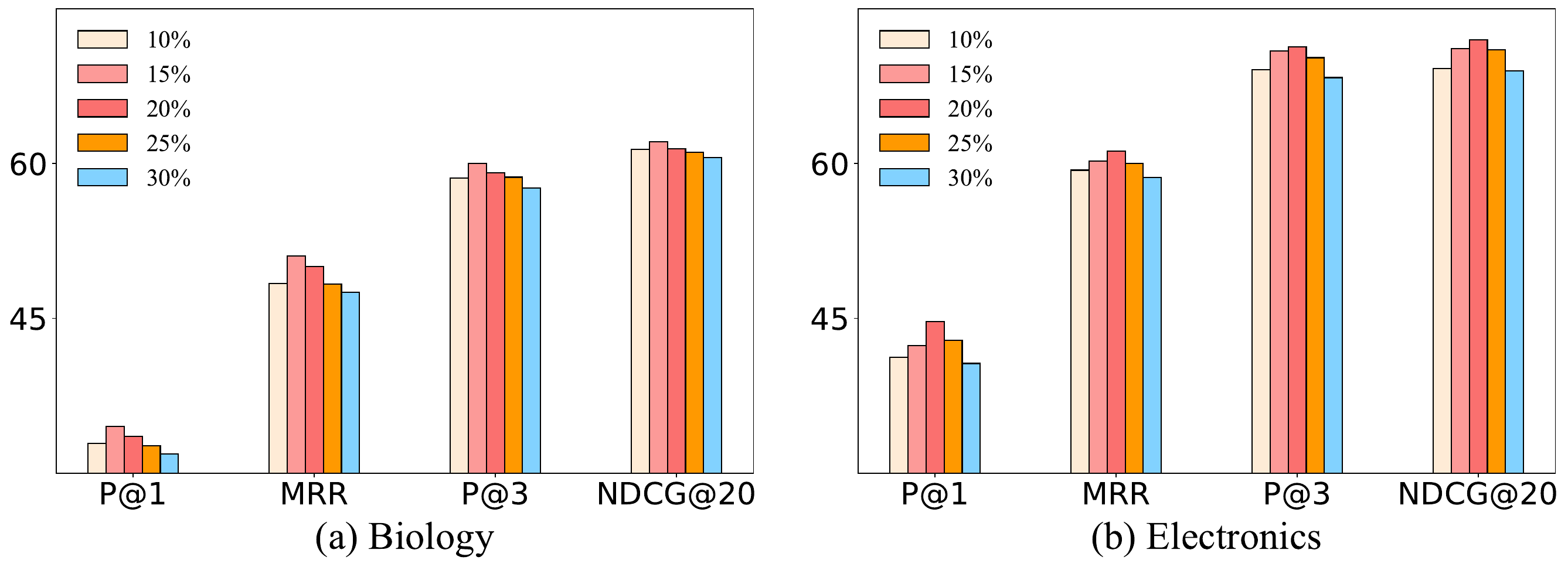}
\caption{Impacts of Different Historical Question Masking Ratios.}
\label{fig:questionratio}
\end{figure*}

\begin{figure*}
\centering
\includegraphics[width=0.8\textwidth]{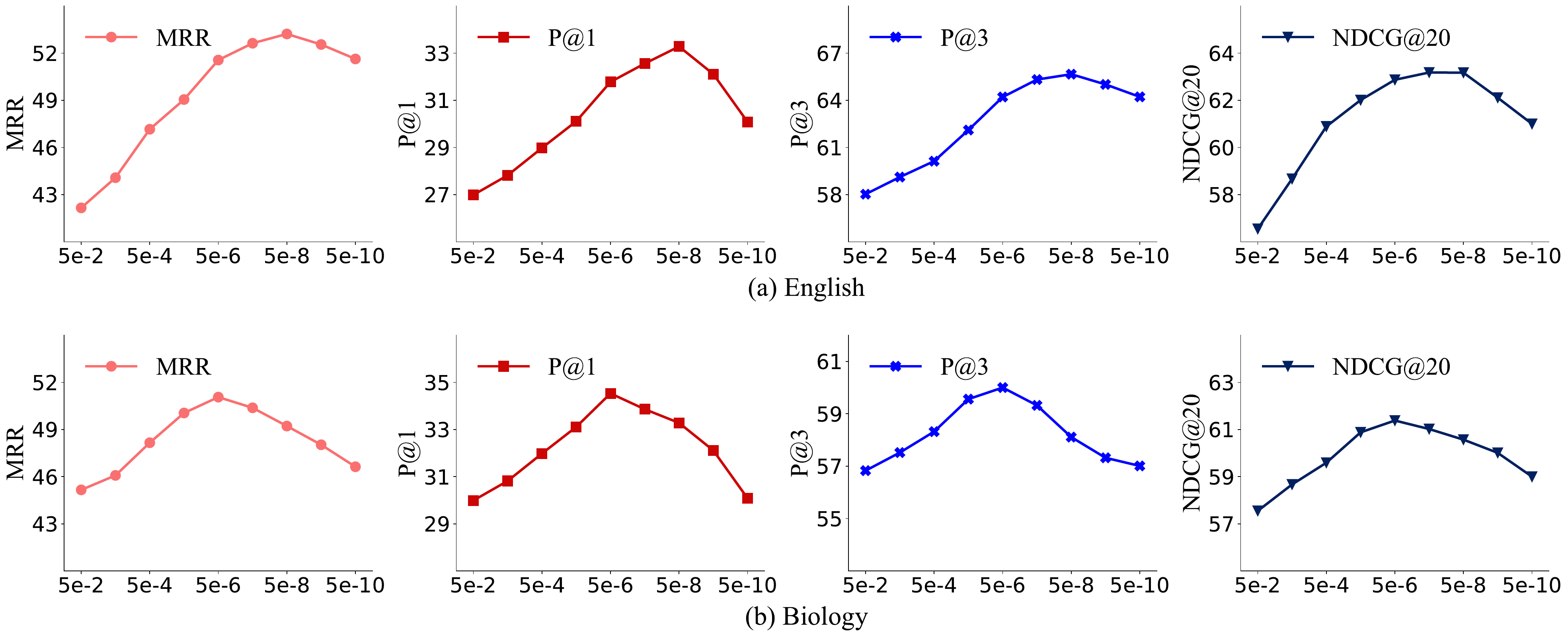}
\caption{Impacts of Different Learning Rates during Fine-tuning.}
\label{fig:learningrate}
\end{figure*}

The results are presented in Fig~\ref{fig:questionratio}. 
From the results, we find:
(1) The ratio of masked questions can impact the model's performance, with an initial increase followed by a decrease as the masking ratio increases.
This is because if the question masking ratio is too small, the role of this pre-training task is weakened.
And the ratio is too large could lead to a weakening of the model’s ability to accurately model experts;
(2) Different datasets have different sensitivity to question mask ratios (e.g., the Biology dataset is $15\%$, the Electronic dataset is $20\%$).
This is because the average number of expert answers in different datasets is also different (e.g., Biology dataset is $18.11$ while Electronic dataset is $33.14$).
Hence, we need to set the question masking ratio carefully.

\subsubsection{Learning Rate in Fine-tuning}

\begin{figure*}
\centering
\includegraphics[width=0.7\textwidth]{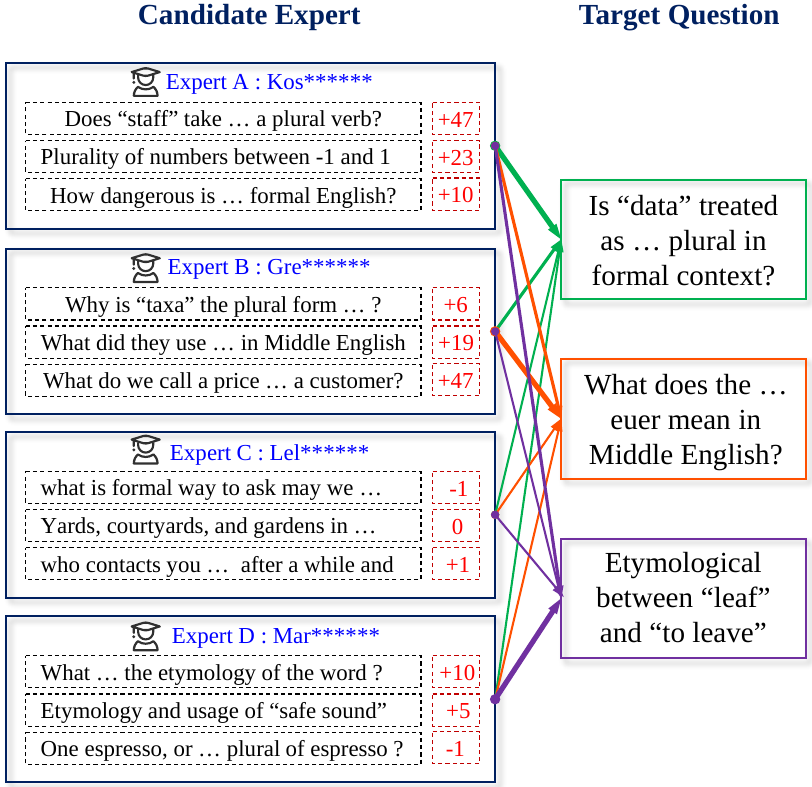}
\caption{Case Study. Visualization of matching scores between experts and questions. The arrows from experts to the same target question have the same color. Meanwhile, the thickness of the arrow represents the matching score (i.e., the thicker the arrow, the higher the matching score).}
\label{fig:casestudy}
\end{figure*}

During fine-tuning, we observe an interesting phenomenon: the learning rate significantly impacts expert finding performance.
As depicted in Fig~\ref{fig:learningrate}, we observe a consistent pattern across different datasets: the model performance initially improves and then gradually declines.
A large learning rate can harm expert representations from pre-training.
Additionally, the inflection point of the curve differs across different datasets.
For instance, the optimal learning rate for the English dataset is 5e-8, while that of the Biology dataset is 5e-6.
We think that this might be due to the size of the dataset.
For larger datasets (e.g., English), a higher learning rate could lead to drastic changes in the pre-trained representations, resulting in degraded expert finding performance. 
Conversely, smaller datasets (e.g., Biology) may require a higher learning rate to achieve optimal performance. 
In summary, we recommend setting the learning rate carefully.

\subsection{Case Study}

In this part, we conduct a case study to better understand what has been learned by the PEPT during pre-training.
Specifically, we randomly select several target questions from the English testing datasets and several pre-trained candidate experts for prediction.
We visualize the final matching score $E_c$ in Fig~\ref{fig:casestudy}, where the thickness of the arrow represents the matching score.
A thicker arrow indicates a higher matching score between questions and candidate experts.

As shown in the Figure, for the target question ``Is data treated as ... plural in formal context'', Expert A obtains the highest matching score and he/she is the ground truth expert. 
This is because Expert A has successfully answered relevant questions in the past and obtained a high vote score (e.g., "Does staff take ... plural verb" with a high vote score of \textbf{+47}).
However, Expert D does not receive a high match score, even though he/she has answered a similar question ("One espresso, or ... plural of espresso").
This is because his/her vote score is unsatisfactory (\textbf{-1}), indicating that he/she may not be suitable for answering this type question.
It is not surprising that Expert C obtains the lowest matching score when matching any question.
This is because Expert C obtained low vote scores in most historical questions, which might not be good for answering these questions.

These case studies demonstrate that our paradigm pre-trains expert representations with both interest and expertise, benefiting expert finding in CQA.

\section{Limitation}

In fact, the vote scores received for each question can roughly measure the answerer's expertise in the question. 
However, because the popularity of questions varies, a low vote score may not necessarily mean that the answer is of poor quality.
For alleviating this, in the future, we will explore the following aspects:
\begin{enumerate}
    \item Categorizing different questions and adjusting the weight of vote scores based on the category popularity. For example, questions in less popular areas usually have fewer votes, so the weight of such question vote scores can be amplified.
    \item Mitigating bias caused by question popularity from a causal perspective. First, we can construct a causal graph to describe the relationship between questions, answers, voting scores, popularity and categories. The relationship might include: category affects question, question affects answer, answer affects voting score, popularity of a question affects the question, and user expertise affects answer, etc. 
    
    In this causal model, the popularity of the question has a indirect impact on the voting score, which may lead to a deviation in the voting scores of questions of different categories. Then we employ inverse probability weighting to adjust the voting score of each question, so that the voting score can more reflect the user's true professional level.
    \item Introducing Large Language Models (LLMs, e.g, ChatGPT) to evaluate the professionalism of the expert’s historical answers because the LLMs has large general knowledge. Specifically, we fine-tune the open source LLMs (e.g., Llama-3) or directly use the closed source LLMs (e.g., ChatGPT) to evaluate the expert expertise. Given one expert, we input historical question and corresponding answer  into the LLMs, and employ the prompt (e.g., You are a CQA expert. I have given you a question and a corresponding answer. Please help me judge the professionalism of the answer (score: 1-5).) to make the model output a professional score. Then we introduce this score into the expert finding model as auxiliary information to help model expert ability.
\end{enumerate}

%% file: 6-conclusion.tex
\section{Conclusion}

In this paper, our focus is on exploring the design of a personalized pre-training language model that is specifically customized for expert finding on CQA websites.
Unlike the previous version ExpertPLM could only capture coarse-grained expert characteristics, we propose a more fine-grained pre-training framework that integrates expert histories, vote scores, and personalized information. 
This framework includes question-level masked language modeling and vote-oriented tasks.
Based on these, the model is enabled to learn personalized expert features that can be transferred to the downstream task of expert finding, which could reduce the dependency on large amounts of labeled data.
Our PEPT model demonstrates superiority expert finding performance on six real-world CQA datasets.
Further ablation studies confirm the individual effectiveness of modeling interest, expertise, and personalization.
Moreover, we conduct an evaluation to assess the performance of our model in zero-shot scenarios, which could validate that the expert personalized information captured by the model during pre-training.